\documentstyle[12pt]{article}
\begin{document}

\title{
%
%
{\normalsize\hfill{UM--P--96/11; RCHEP 96/01}}\vspace{4pc}\linebreak
\large\bf Quaternionic Formulation of the Exact Parity Model}

\author{\normalsize S.\ P.\ Brumby, R.\ Foot and R.\ R.\ Volkas\\
\\
\normalsize\it Research Centre for High Energy Physics, \\
\normalsize\it School of Physics, University of Melbourne, \\
\normalsize\it Parkville, Victoria 3052, Australia}

\date{}

\maketitle

\begin{abstract}
The exact parity model (EPM) is a simple extension of the Standard
Model which reinstates parity invariance as an unbroken symmetry of
nature. The mirror matter sector of the model can interact with 
ordinary matter through gauge boson mixing, Higgs boson mixing and, 
if neutrinos are massive, through neutrino mixing. The last effect 
has experimental support through the observed solar and atmospheric
neutrino anomalies. In this paper we show that the exact parity model
can be formulated in a quaternionic framework. This suggests that the
idea of mirror matter and exact parity may have profound implications
for the mathematical formulation of quantum theory.
\end{abstract}

\pagebreak

\section{Introduction}

The exact parity model (EPM) is a simple extension of the Standard
Model which reinstates parity invariance as an exact symmetry of
nature \cite{lee,flv}. 
The mirror matter sector of the model can interact with ordinary 
matter through gauge boson mixing, Higgs boson mixing and, if
neutrinos are massive, through neutrino mixing. The last effect has
experimental support through the observed solar and atmospheric
neutrino anomalies \cite{nu}. 
In this paper we show that the exact parity model
can be formulated in a quaternionic framework. This suggests that the
idea of mirror matter and exact parity may have profound implications
for the mathematical formulation of quantum theory.

The Standard Model (SM), and any extension thereof, can be augmented
with a mirror matter sector in order to reinstate parity invariance as
an exact symmetry (neither explicitly nor spontaneously broken). Take
the SM as a concrete example. The gauge group $G_{SM} = $SU(3)$_c
\otimes$SU(2)$_L \otimes$U(1)$_Y$ of the SM is first extended by
postulating that the gauge theory of the world is $G_{SM} \times
G'_{SM}$. The ordinary fermion and Higgs fields of the SM are placed
into the usual representations of $G_{SM}$ and are taken to be
singlets under $G'_{SM}$. One then introduces {\it mirror matter} as
fields that are singlets under $G_{SM}$ but have the standard
assignments under $G'_{SM}$. In order to make the theory parity
symmetric, the mirror fermion fields are taken to have the opposite
chirality to the ordinary fermion fields, and an exact parity symmetry
which interchanges ordinary and mirror fields is introduced. It is
simple to show that this non-standard parity symmetry is not
spontaneously broken by the Higgs potential for a large range of
parameters \cite{flv}. 

Ordinary and mirror matter share common gravitational
interactions, which makes mirror matter interesting for cosmology
\cite{cosmo}.
Perhaps of even more interest is that ordinary and mirror matter will
generically also interact through non-gravitational interactions. In
the case of the exact parity extension of the SM these
non-gravitational interactions are induced by photon--mirror-photon,
$Z$--mirror-$Z$ and Higgs mixing. Photon mixing causes mirror
particles to have ordinary electric charge. Since mirror matter has
yet to be observed electromagnetically, an upper bound on the photon
mixing parameter can be derived \cite{glashow}. 
This phenomenological constraint
forces the ordinary electric charges of mirror particles to be very
small, or ``minicharged'' using terminology in the literature. A very
interesting and characteristic prediction of the EPM is thus that
there will exist mirror leptons and quarks which are mass degenerate
with ordinary leptons and quarks, and which have exactly the same
electric charge {\it ratios} as the ordinary fermions. The $Z$ boson
mixing phenomenon is constrained to be very small because it is
controlled by the same parameter that induces photon mixing. This
effect is so small that it is phenomenologically uninteresting. Higgs
boson mixing also yields a characteristic prediction: because mass 
eigenstates must also be parity eigenstates, the physical
neutral Higgs boson will be maximally mixed with its mirror partner.
Thus each mass eigenstate Higgs boson will decay half of the time into
ordinary matter and the other half into mirror matter
\cite{flv,solarnu}. This prediction is testable provided that the
Higgs boson mixing parameter is large enough.\footnote{Actually, there
will be quite strong constraints on the Higgs boson mixing parameter
if one demands that the mirror sector does not come into equilibrium
with the ordinary sector in the early universe (in the context of the
Hot Big Bang model). It may be that these constraints imply that 
the physical Higgs boson mass eigenstates cannot be
resolved in experiment.} 

If neutrinos are massive, then ordinary and mirror neutrinos will
generically mix. The exact parity symmetry of the theory forces the
mixing angle between and ordinary neutrino and its mirror partner to
be maximal just as for Higgs boson mixing.
In the absence of large inter-generational neutrino mixing (as suggested 
by the observed form of the Kobayashi-Maskawa matrix), this leads 
for a large range of parameters to the
observationally supported prediction that the flux of both solar and
atmospheric neutrinos should be half of the standard expectation. This
scenario will be tested further by the forthcoming SNO,
SuperKamiokande and Borexino experiments, as well as by the
continuation of the GALLEX and SAGE experiments.

The EPM is thus of great interest for both theoretical and
phenomenological reasons. In this paper, we wish to explore a possible
connection between the EPM and quaternions. Our starting point is the
following observation: In the EPM, every particle has a distinct
mirror partner. In other words, the number of degrees of freedom is
exactly doubled. Our idea is that this doubling of degrees of freedom
may be related to the doubling of degrees of freedom concomitant with
the extension of complex quantum mechanics to quaternionic quantum mechanics.
Therefore, we will introduce quaternionic fields and construct our theory
so that the extra fields correspond to the particles in the  mirror sector
of the EPM.

While the fields in our theory will be quaternionic, we will consider 
the underlying quantal structure to be that of ordinary complex
quantum mechanics.  That is, we will ascribe
special significance to a particular pure imaginary unit quaternion
and identify it with the $i$ of normal (complex) quantum mechanics.

Although there are some interesting ideas in the literature about how 
one may write down a genuinely quaternionic quantum theory
\cite{FJSS,Horw,Adle,Joshi}, there is no empirical evidence which would
allow the selection of the best candidate.
Furthermore, it is possible to conceive of mechanisms by which quaternionic 
processes operating in the high energy (or short distance) regime become
negligible in a low energy (or spatially asymptotic) limit \cite{Adle}
retrieving standard complex dynamics but with possible extra (remnant,
physical) degrees of freedom.  
We will therefore take it as premature to try
to develop the EPM on the basis of a full quaternionic quantal structure,
although we do not preclude this as a future development. Indeed we
hope that the present work may act as a spur for such developments.
For the moment we will take the first step: to rewrite the usual
Lagrangian of the EPM using quaternionic fields. The quaternionic
parts of each individual term in the Lagrangian will however exactly
cancel with the quaternionic parts of another term in the Lagrangian.
In this sense, we will not be doing anything radical. However, we will
set the stage for a fully quaternionic treatment. We hope that you
will agree with us that even our limited exercise leads to interesting
mathematics, and that interesting physics may develop from it in the
future. It is certainly non-trivial that the EPM can be rewritten in
this particular manner.

Before we begin, we need to state our notational conventions. 
We choose the metric 
tensor to be given by $g^{\mu\nu} = {\rm diag}(-1,1,1,1)$, with the
Dirac matrices obeying the usual relations
$\{\gamma^{\mu},\gamma^{\nu}\} = 2g^{\mu\nu}$. We will work in the
Majorana representation for the Dirac matrices. With the above metric,
each of the $\gamma^{\mu}$ is a real matrix in the Majorana
representation. Further, $\gamma^{0T} = -
\gamma^0$ while $\gamma^{iT} = \gamma^{i}$. The matrix $\gamma_5
\equiv i\gamma^0\gamma^1\gamma^2\gamma^3$ is pure imaginary. The
fact that the $\gamma^{\mu}$'s are real while $\gamma_5$ is imaginary
will prove extremely convenient when we need to commute quaternion
units with Dirac matrices.

\section{Lagrangian for a free fermion}

Consider a fermionic field $\Psi$ which is
quaternion valued. We can write $\Psi$ as
\begin{equation}
\Psi = \psi_0 + i \psi_1 + j \psi_2 + k \psi_3,
\end{equation}
where $\psi_0,...,\psi_3$ are real-valued and $i$, $j$ and $k$ are the
quaternion units. They obey the quaternionic multiplication rules,
\begin{equation}
i^2 = j^2 = k^2 = -1\quad {\rm and}\quad ijk = -1.\label{ijk}
\end{equation}
Note that the quaternions are not commutative. 

As discussed in the Introduction, we will not be pursuing quaternionic
quantum mechanics as such.  Instead we choose $i$ to play the usual role
of $\sqrt{-1}$ in quantum mechanics.  That is, $i$ commutes with all
the operators corresponding to generalised co-ordinates and their
conjugate momenta, and appears in all the fundamental commutators.
We anticipate that complex analytical tools (such as the Fourier transform)
can be extended to our quaternionic Hilbert space, with bells and whistles
attached, but our present work will be limited to recasting the 
standard Lagrangian of the (complex) EPM into an exactly equivalent form
written in terms of quaternions, and we shall avoid attempting to formulate
a full theory of quantised, quaternionic fields\footnote{%
There is no simple quaternionic generalisation of Fourier transform
technology \cite{Sud:79}.
Adler \cite{Adle} has proposed using (real) Fourier-sine and
-cosine transforms.}.

Having singled out $i$ to play a special role, we take advantage of the 
symplectic representation of $\Psi$, whereby
\begin{equation}
\Psi = \psi + j \psi'.
\end{equation}
The fields $\psi \equiv \psi_0 + i \psi_1$ and $\psi' \equiv \psi_2 -
i \psi_3$ are ``complex''-valued fields (i.e., complex with respect to 
the sub-algebra generated by $1$ and $i$).

We will first rewrite the Lagrangian for a free Dirac fermion in terms
of $\Psi$. It is given by ${\cal L}_{K}$ where
\begin{equation}
{\cal L}_K = \frac{1}{2} ( i \overline{\Psi} \partial\!\!\!/ L \Psi +
\overline{\Psi} \partial\!\!\!/ L \Psi i ),
\end{equation}
where $\partial\!\!\!/ \equiv \gamma^{\mu}\partial_{\mu}$, and $L$ is
the usual left-handed chiral projection operator $(1 - \gamma_5)/2$. 
Note that the second term above is the Hermitian conjugate of the 
first term, and it follows from the algebraic relations amongst the
quaternionic units, Eqs~(\ref{ijk}), that we have completely removed the
$j$ and $k$ dependence of the Lagrangian.

We can equally well write this Lagrangian as
\begin{equation}
{\cal L}_K = \frac{1}{2} ( i \overline{\Psi}_L \partial\!\!\!/ \Psi_L +
\overline{\Psi}_L \partial\!\!\!/ \Psi_L i ),
\end{equation}
where $\Psi_L \equiv L \Psi$ is a chiral quaternionic fermion field.

The relevance of quaternions to the notion of exact parity can be
simply illustrated by examining $\Psi_L$ more closely. We know that
\begin{equation}
\Psi_L \equiv L \Psi = L \psi + L j \psi' = L \psi + j R \psi',
\end{equation}
where $R \equiv (1 + \gamma_5)/2$ is the usual right-handed chiral
projection operator. Since $\gamma_5$ in the adopted notational scheme
is pure imaginary, it anticommutes with the quaternion unit $j$. We
have therefore established that the left-handed chiral quaternionic
spinor $\Psi_L$ is a symplectic linear combination of an ordinary
left-handed Weyl spinor $\psi_L \equiv L\psi$ and an ordinary 
right-handed Weyl spinor $\psi'_R \equiv R \psi'$.   

Under
the parity transformation $x^{\mu} \to x_{\mu}$, $\psi_L \to -\gamma^0
\psi'_R$ and $\psi'_R \to \gamma^0 \psi_L$ (the minus sign in
this transformation ensures that it is a $Z_2$ transformation). 
There are two useful ways
to establish this. First one can simply substitute for $\Psi$ in terms
of the complex-valued fields $\psi$ and $\psi'$ in ${\cal L}_K$. The
result is
\begin{equation}
{\cal L}_K = i \overline{\psi} \partial\!\!\!/ L \psi + i
\overline{\psi}' \partial\!\!\!/ R \psi'
=  i \overline{\psi}_L \partial\!\!\!/ \psi_L + i \overline{\psi}'_R
\partial\!\!\!/ \psi'_R 
\label{LKcomplex}
\end{equation}
which is manifestly parity invariant in the usual way.
There is also an elegant quaternionic derivation.
The parity transformation written in terms of the quaternionic field
$\Psi$ is
\begin{equation}
x^{\mu} \to -x_{\mu}\quad {\rm and}\quad \Psi_L \to j \gamma^0 \Psi_L.
\label{parity}
\end{equation}
It is easy to show that ${\cal L}_K$ is invariant under this
transformation. If $\Psi_L \to j \gamma^0 \Psi_L$ is written in terms of
the symplectic components one obtains $\psi_L \to -\gamma^0 \psi'_R$
and $\psi'_R \to \gamma^0 \psi_L$ which is just the usual parity
transformation. 

The simple calculation above establishes an interesting connection
between parity symmetry and quaternions. However, as is well known
there are many ways to extend the SM Lagrangian into a parity symmetric
form. One way is through the use of the idea of ``left-right
symmetry'' \cite{lrs} 
and another is the mirror matter formulation we are
interested in here. The two are distinguished for three reasons. First,
for phenomenological reasons left-right symmetry can only be a 
symmetry of the Lagrangian, not of the vacuum. By contrast, parity
invariance as implemented through mirror matter can be a symmetry of
both the Lagrangian and the vacuum. Second, the left-right symmetric
model sees parity-partner fields sharing common
SU(3)$_c\otimes$U(1)$_{B-L}$ gauge interactions, while mirror matter
is constructed to be neutral under all $G_{SM}$ gauge forces (modulo
the neutral gauge boson mixing effects mentioned earlier). 
Third, in left-right symmetric models the chiral partners forming a
Dirac fermion are also parity partners, whereas the parity partners in
the EPM are {\it not} also mass partners.
Does the quaternionic formulation of parity distinguish between 
these two scenarios?

\section{Fermion mass}

We will begin to answer this question by examining the third of the
distinguishing features listed above. Let us examine the varieties of
fermion mass possible in the quaternionic framework. 

One may construct meaningful mass terms purely from the chiral
quaternionic field $\Psi_L$. These are in a sense the simplest mass
terms one can have, because they do not involve the introduction of
additional degrees of freedom through a right-handed quaternionic
partner field $\Psi_R$. Consider the term
\begin{equation}
B_1 \equiv \overline{\Psi}_L j \Psi_L.
\end{equation}
This term is Lorentz invariant and nonzero, because the right chiral
projector $R$ that multiplies $\overline{\Psi}$ from the right turns
into $L$ after commutation through $j$: 
$B_1 = \overline{\Psi}jL\Psi$. Furthermore, it is Hermitian as
can be seen from the following simple computation:
\begin{equation}
\left(\Psi_L^{\dagger}\gamma^0 j \Psi_L \right)^{\dagger}
= \Psi_L^{\dagger} (-j) \gamma^{0\dagger} \Psi_L
= \Psi_L^{\dagger}\gamma^0 j \Psi_L,
\end{equation}
where in the last step we have used the fact that $\gamma^{0\dagger} =
- \gamma^0$ in our notation and that $j$ commutes with the real matrix
$\gamma^0$. Finally, $B_1$ is parity symmetric under
Eq.~(\ref{parity}). In terms of symplectic components,
\begin{equation}
B_1 = \overline{\psi}'_R \psi_L - \overline{\psi}_L \psi'_R +
\overline{\psi}_L j \psi_L + \overline{\psi}'_R j \psi'_R,
\end{equation}
where the minus sign between the first two terms combines with the
phase factor of $-1$ obtained from Eq.~(\ref{parity}) to render the first two
terms a parity symmetric pair. The last two terms are also obviously
parity symmetric. However, the last two terms are $j$-dependent and
our programme requires that we expunge them. In a true quaternionic
quantum mechanics it may be interesting to ascribe physical significance
to such peculiar $j$-dependent mass terms, but for our present
purposes we must render them unphysical. This is achieved by a brute
force cancellation. We simply require that the physically allowed mass
term modelled on $B_1$ be given by ${\cal B}_1$ \footnote{%
 From a mathematical point of view, we are projecting the mass term 
into the $i$-complex sub-algebra of the quaternions. The properties of 
quaternionic mappings of the type $Q\ \rightarrow\ (Q - qQq)/2$ are 
investigated by de Leo, {\em et al\/} \cite{deLeo}.}
where
\begin{equation}
{\cal B}_1 \equiv \frac{1}{2}(B_1 - i B_1 i).
\end{equation}
The relation $-iji = -j$ then ensures that all $j$-dependent pieces of
$B_1$ are systematically cancelled by the second term. In symplectic
components we then simply have that
\begin{equation}
{\cal B}_1 = \overline{\psi}'_R \psi_L - \overline{\psi}_L \psi'_R.
\end{equation}
Note that this term identifies parity partners as also mass partners.
It is therefore suitable for use in a possible quaternionic
formulation of the usual left-right symmetric model, but not for the
EPM. It will ultimately be gauge invariance that distinguishes between
left-right symmetry and the EPM. The above mass term will be
chosen to be gauge invariant in the left-right symmetric case (after
electroweak symmetry breaking), but it will not be gauge invariant in
the EPM.

There are several other suitable fermion bilinears. We will discuss
them all below, but we will not provide the (simple) derivations of
their stated properties as we did in the warm up example above.

Our second bilinear is
\begin{equation}
{\cal B}_2 \equiv \frac{1}{2} (B_2 - i B_2 i),
\end{equation}
where
\begin{equation}
B_2 \equiv \overline{\Psi}_L k \Psi_L.
\end{equation}
This is simply the parity-odd version of ${\cal B}_1$. A linear
combination of ${\cal B}_1$ and ${\cal B}_2$ with arbitrary real
coefficients will then yield the most general hermitian Dirac mass
term connecting the symplectic components.

The next class of bilinears provide Majorana masses. In the Majorana
representation of the Dirac matrices, the charge conjugation matrix
${\cal C}$ is simply given by
\begin{equation}
{\cal C} = -i \gamma^0.
\end{equation}
Using this matrix we can write down the charge-conjugate of $\Psi_L$,
given by
\begin{equation}
(\Psi_L)^c \equiv {\cal C} \left( \overline{\Psi}_L \right)^T =
(\psi_L)^c + j (\psi'_R)^c.
\end{equation}
We therefore see that the $(\Psi_L)^c$ has opposite chirality to
$\Psi_L$ as usual. 

The parity even or symmetric Majorana bilinear is given by
\begin{equation}
{\cal B}_3 \equiv \frac{1}{2} (B_3 - i B_3 i) + {\rm H.c.}
\end{equation}
where
\begin{equation}
B_3 \equiv \overline{\Psi}_L (\Psi_L)^c.
\end{equation}
The parity odd term is
\begin{equation}
{\cal B}_4 \equiv \frac{1}{2} (B_4 - i B_4 i) + {\rm H.c.}
\end{equation}
where
\begin{equation}
B_4 \equiv i \overline{\Psi}_L i (\Psi_L)^c.
\end{equation}
A general Majorana mass is then an arbitrary linear combination of the
two terms above, giving different Majorana masses to $\psi_L$ and
$\psi'_R$.

This concludes the catalogue of all acceptable mass terms that can be
formed using $\Psi_L$ only. We have obviously covered all of the
possibilities offered by the ordinary symplectic components $\psi_L$
and $\psi'_R$: all varieties of acceptable mass involving these
fields are included above. We can therefore be sure that we have left
no independent possibilities at the quaternionic level unaccounted for.

We now introduce a right-handed quaternionic spinor 
$\Psi_R = \psi_R + j\psi'_L$ in addition to $\Psi_L$. The complete
list of Lorentz-invariant, Hermitian and nonzero terms that do not
involve charge conjugation is:
\begin{eqnarray}
B_5 & \equiv & i \overline{\Psi}_L \Psi_R + \overline{\Psi}_R
\Psi_L i,\\
B_6 & \equiv & \overline{\Psi}_L i \Psi_R + \overline{\Psi}_R
i \Psi_L,\\
B_7 & \equiv & j  \overline{\Psi}_L \Psi_R + \overline{\Psi}_R
\Psi_L j,\\
B_8 & \equiv & \overline{\Psi}_L \Psi_R j + j \overline{\Psi}_R
\Psi_L,\\
B_9 & \equiv &  k  \overline{\Psi}_L \Psi_R + \overline{\Psi}_R
\Psi_L k,\\
B_{10} & \equiv & \overline{\Psi}_L \Psi_R k + k \overline{\Psi}_R
\Psi_L.
\end{eqnarray}
Other possibilities such as $\overline{\Psi}_L j \Psi_R +
\overline{\Psi}_R j \Psi_L$ are identically equal to zero. These terms
are then made $j$-independent using the trick introduced earlier. One
then obtains the allowed set
\begin{equation}
{\cal B}_{5-10} \equiv \frac{1}{2} ( B_{5-10} - i B_{5-10} i),
\end{equation}
of standard mass terms rewritten in terms of quaternionic fields.

In symplectic components, 
\begin{equation}
{\cal B}_5 = i (\overline{\psi}_R \psi_L +
\overline{\psi}'_L \psi'_R + \overline{\psi}_L \psi_R +
\overline{\psi}'_R \psi'_L).
\end{equation}
These terms produce a degenerate pair of Dirac fermions
given by $\psi \equiv \psi_L + \psi_R$ and $\psi' \equiv \psi'_L +
\psi'_R$. (The unusual overall factor of $i$ is simply due to our
chosen notational scheme.) Furthermore, using Eq.~(\ref{parity}) one
sees that ${\cal B}_5$ is parity invariant.
The interesting observation is that the parity partner of, say,
$\psi_L$ is $\psi'_R$ rather than its mass partner $\psi_R$. This type
of mass term is precisely the one we will need to use to construct the
quaternionic reformulation of the EPM.\footnote{One can define another
version of the parity transformation as $\Psi_{L,R} \to i\gamma^0
\Psi_{R,L}$ under which ${\cal B}_5$ is also invariant. This is
actually another way of defining standard parity, because the parity
partners $\psi_{L,R}$ and $\psi'_{L,R}$ are again also mass partners.
However this definition of standard parity is only possible if the
number of degrees of freedom is doubled (one has $\psi'$ as well as
$\psi$). Clearly ${\cal B}_5$ is invariant under both standard parity
and the mirror matter version of parity. With it one could construct a
left-right symmetric model augmented by a mirror matter sector, a
perfectly consistent and reasonable model which however we will not
pursue any further. Who needs a broken mirror in addition to a good
mirror when just one good mirror may be enough to reflect reality?}

The term ${\cal B}_6$ is just the parity-odd version of ${\cal B}_5$.
In terms of symplectic components it is given by
\begin{equation}
{\cal B}_6 = i (\overline{\psi}_R \psi_L - 
\overline{\psi}'_L \psi'_R + \overline{\psi}_L \psi_R -
\overline{\psi}'_R \psi'_L).
\end{equation}
If parity symmetry was not imposed on the Lagrangian, then the most
general mass Lagrangian would be $m{\cal B}_5 + m' {\cal B}_6$. The
ordinary complex Dirac
fermions $\psi$ and $\psi'$ would then no longer be mass degenerate.

The term ${\cal B}_7$ is an interesting case. Focus first of all on
just the $j\overline{\Psi}_L \Psi_R$ part. A simple computation shows
that
\begin{equation}
j\overline{\Psi}_L \Psi_R = j \overline{\psi}_L j \psi'_L - 
j \overline{\psi}_R j \psi'_R + j (\overline{\psi}_L \psi_R +
\overline{\psi}'_R \psi'_L).
\end{equation}
The latter two terms have a nonzero quaternionic piece which will be
cancelled by other terms in ${\cal B}_7$, so we can ignore them.
To interpret the remaining terms, it is
necessary to first remove the $j$'s by bringing them to adjacent
positions and then using $j^2 = -1$. One finds that
\begin{equation}
j \overline{\psi}_L j \psi'_L = - \psi^T_L \gamma^0 \psi'_L 
= i \overline{(\psi_L)^c} \psi'_L.
\end{equation}
The term ${\cal B}_7$ therefore describes off-diagonal Majorana masses
linking $\psi$ with $\psi'$.

The other terms ${\cal B}_{8,9,10}$ just describe other possibilities
for these off-diagonal Majorana masses. For instance ${\cal B}_8$ is
the parity partner of ${\cal B}_7$. Similarly ${\cal B}_9$ and ${\cal
B}_{10}$ are parity partners. Adding ${\cal B}_{7}-{\cal B}_{10}$ with
arbitrary coefficients produces the most general allowed off-diagonal
Majorana masses.

The ingredients $\psi_{L,R}$ and $\psi'_{L,R}$ can only produce the
Dirac and Majorana masses covered by ${\cal B}_5-{\cal B}_{10}$ above.
We can therefore be sure that all meaningful independent mass terms
that one can write using quaternionic fields are included in this
list. Any other candidates one could think of, such as 
$\overline{\Psi_L} j (\Psi_R)^c$ for example, cannot be independent of
the terms already considered.

To summarise this section then, we have shown above that both
left-right symmetric parity and mirror matter parity can be
accommodated by mass terms formed out of quaternionic spinors. Parity
asymmetric masses can also be included. If ${\cal B}_5$ is chosen
as the preferred mass, then 
the symplectic components of the fermion fields
form the ordinary and mirror sectors. It will ultimately be gauge
invariance that distinguishes between these cases. It should also be
clear that the symplectic components of a quaternionic spinor can be
constructed to be fermions that interchange under {\it any} $Z_2$ discrete
symmetry. As well as parity, one can also incorporate discrete
symmetries such as charge conjugation or a $Z_2$ horizontal symmetry.
We will not explicitly consider these possibilities here, but the
reader should be aware of them.

If parity symmetry (or any other $Z_2$ symmetry) is not
imposed, then the fermions in the theory still form two sectors, but
there is no symmetry between them and thus no mass degeneracy. In the
context of the SM, the parity non-invariant quaternionic formulation
can for instance lead to a $G_{SM} \otimes G'_{SM}$ 
gauge theory with a completely
independent set of Yukawa coupling constants and other parameters for
each of the sectors. We will henceforth have to explicitly 
impose parity invariance on the theory. 

\section{Global Abelian transformations}

Let us now consider symmetry transformations on the quaternionic fermion
field $\Psi_L$. We begin with global Abelian transformations. Because
the fermion field $\Psi_L$ is quaternionic (and hence
non-commutative), there are two possible
phase invariances: multiplication on the left and multiplication on
the right. If we do these simultaneously the most general
transformation is
\begin{equation}
\Psi_L \to e^{i\alpha} \Psi_L e^{i\beta},
\end{equation}
where $\alpha$ and $\beta$ are independent of each other.\footnote{Note
that we do not consider transformations such as $\Psi_L \to
e^{j\alpha}\Psi_L e^{j\beta}$ because they do not commute with the
Lorentz group.} In
symplectic components this transformation is
\begin{equation}
\psi_L + j \psi'_R \to e^{i(\alpha + \beta)} \psi_L + j e^{i(-\alpha +
\beta)} \psi'_R,
\end{equation}
where $e^{i\alpha}$ has been commuted through $j$. The complex fields
themselves therefore simply undergo two independent Abelian
transformations:
\begin{equation}
\psi_L \to e^{i(\alpha + \beta)} \psi_L\quad {\rm and}\quad \psi'_R 
\to e^{i(-\alpha + \beta)} \psi'_R.
\label{globalAb}
\end{equation}
This is of course exactly what we need in order to have ordinary and
mirror matter transform under independent groups (the two independent
U(1)$_Y$ factors of $G_{SM}\otimes G'_{SM}$ for instance).

Furthermore, it is clear from Eqs.~(\ref{globalAb}) and
(\ref{LKcomplex}) that ${\cal L}_K$
is invariant under these global Abelian transformations. A problem
that will continue to arise in the ensuing discussion is how best to
establish the invariance of a candidate Lagrangian that is written in
terms of quaternionic fields. The observation we have just made about
the invariance of ${\cal L}_K$ under global Abelian transformations
relied on us decomposing the quaternionic fields into their complex
components. Although it is rather helpful to do this, it would be more
elegant if an explicitly quaternionic way could be found to prove
invariance. We will make some progress towards this goal, although we
will see that in many ways it will remain easier to prove invariance
at the complex rather than at the quaternionic level. We regard the
use of complex decomposition as an unfortunate device, and it is to be
hoped that a transparent quaternionic method eventually gets
constructed.

In order to facilitate all further discussion of this and other issues,
we now introduce a simple but powerful idempotent operator technology. 
Consider operators $\hat{C}_+$ and $\hat{C}_-$ defined by \cite{deLeo}
\begin{equation}
\hat{C}_+ \Psi_L \equiv -\frac{i}{2}\{i, \Psi_L\} = \psi_L\quad 
{\rm and}\quad \hat{C}_- \Psi_L \equiv -\frac{i}{2} [i, \Psi_L] =
j\psi'_R.
\end{equation}
These operators obey the relations
\[
[i, \hat{C}_{\pm}] = 0\,,
\]
\begin{equation}
\hat{C}_+^2 = \hat{C}_+,\quad \hat{C}_-^2 = \hat{C}_-,
\label{ops}
\end{equation}
\[
\hat{C}_+\hat{C}_- = \hat{C}_-\hat{C}_+ = 0,\quad \hat{C}_+ + \hat{C}_- = 1.
\]
They therefore project out the symplectic components from $\Psi_L$,
with $\hat{C}_+$ projecting out $\psi_L$ and $\hat{C}_-$ projecting
out $j\psi'_R$. The global U(1)$\otimes$U(1) transformation can
therefore be rewritten as
\begin{equation}
\Psi_L \to e^{i(\alpha + \beta)} \hat{C}_+ \Psi_L + e^{i(\alpha -
\beta)} \hat{C}_- \Psi_L.
\label{u1xu1}
\end{equation}
Two special cases are of note. If $\beta = 0$, then the transformation
is, trivially,
\begin{equation}
\Psi_L \to e^{i\alpha}(\hat{C}_+ + \hat{C}_-)\Psi_L =
e^{i\alpha}\Psi_L,
\end{equation}
as required. If $\alpha = 0$, the transformation is
\begin{equation}
\Psi_L \to (e^{i\beta} \hat{C}_+ + e^{-i\beta} \hat{C}_-)\Psi_L,
\label{beta1}
\end{equation}
which can be more neatly written as
\begin{equation}
\Psi_L \to e^{i\beta(\hat{C}_+ - \hat{C}_-)} \Psi_L = \Psi_L
e^{i\beta}
\label{beta2}
\end{equation}
using Eqs~(\ref{ops}). The idempotent operators can therefore be used
to rewrite multiplication on the right by a phase as multiplication on
the left by an operator valued phase. This will be useful in writing
non-Abelian transformations. 

An instructive digression is warranted at this point. The idempotent
operator formalism just discussed is closely analogous to the case of
Abelian chiral transformations. Consider a standard massless
Dirac field $\chi
= \chi_L + \chi_R$ where the chiral projection operators $L$ and $R$
are analogous to $\hat{C}_+$ and $\hat{C}_-$. We know that the
Lagrangian of a free massless Dirac field is invariant under
independent phase rotations of its left and right handed components.
This can written as $\chi \to \exp[i(\alpha + \beta)] L \chi 
+ \exp[i(-\alpha + \beta)] R \chi$ which is similar to
Eq.~(\ref{u1xu1}). If $\beta = 0$ then $\chi \to \exp(i\alpha)(L +
R)\chi = \exp(i\alpha)\chi$. If $\alpha = 0$ then 
\begin{equation}
\chi \to (e^{i\beta} L + e^{-i\beta} R) \chi = e^{i\beta(L - R)}\chi =
e^{-i\beta\gamma_5}\chi.
\end{equation}
One useful perspective is that 
the operators $\hat{C}_+$ and $\hat{C}_-$ project out the parity partners
that form a chiral quaternionic spinor, while $L$ and $R$ project out 
the parity
partners that form an ordinary complex Dirac fermion. Another
perspective is that $\hat{C}_+$ and $\hat{C}_-$ project out
irreducible representations of the Lorentz group from the reducible
representation under which $\Psi_L$ transforms, while $L$ and $R$
project out the irreducible parts of $\chi$.

We will now revisit the proof of the invariance of ${\cal L}_K$ under
global Abelian transformations, and use the idempotent operators to
construct a quaternionic derivation. We first note that multiplication
of $\Psi_L$ on the left by $e^{i\alpha}$ poses no problems; invariance
at the quaternionic level is manifest. Our focus is thus necessarily
on multiplication by $e^{i\beta}$ on the right as per Eq.~(\ref{beta2}).
By Hermitian conjugation Eq.~(\ref{beta2}) implies that 
\begin{equation}
\overline{\Psi}_L \to \overline{\Psi}_L e^{-i\beta(\hat{C}_+^{\dagger}
- \hat{C}_-^{\dagger})},
\end{equation}
where the action of $\hat{C}_{\pm}^{\dagger}$ on $\overline{\Psi}_L$
follows immediately from the definition of the operators:
\begin{equation}
\overline{\Psi}_L \hat{C}_+^{\dagger} = \overline{\psi}_L\quad {\rm
and}\quad \overline{\Psi}_L \hat{C}_-^{\dagger} = - \psi'_R j.
\end{equation}
The technical obstacle we face is what to do with the statement that
\begin{equation}
\overline{\Psi}_L \partial\!\!\!/ \Psi_L \to \overline{\Psi}_L
e^{-i\beta(\hat{C}_+^{\dagger} - \hat{C}_-^{\dagger})}
e^{i\beta(\hat{C}_+ - \hat{C}_-)} \partial\!\!\!/\Psi_L, 
\end{equation}
where it is not immediately clear how to combine the two exponentials
in order to establish gauge invariance. Note in particular that
$\hat{C}_{\pm}$ are {\it not} Hermitian (and are thus not projection
operators in the strict mathematical sense).

However, using Eq.~(\ref{beta1}) we easily see that
\begin{equation}
e^{-i\beta(\hat{C}_+^{\dagger} - \hat{C}_-^{\dagger})}
e^{i\beta(\hat{C}_+ - \hat{C}_-)} =
\hat{C}_+^{\dagger} \hat{C}_+ + \hat{C}_-^{\dagger} \hat{C}_- 
+ e^{-2i\beta}\hat{C}_+^{\dagger} \hat{C}_- + e^{2i\beta}
\hat{C}_-^{\dagger} \hat{C}_+.
\label{identity}
\end{equation}
The first two terms are $\beta$-independent and thus constitute the
invariant part of the product of the exponentials. The last two terms
are however $\beta$-dependent and therefore {\it not}
invariant. Notice, though, that the last two terms have products of
$\hat{C}_+$ with $\hat{C}_-$ (up to Hermitian conjugation) and so they
give rise to $j$-dependent terms when inserted between the quaternionic
spinors in ${\cal L}_K$. Since all $j$-dependence cancels from ${\cal
L}_K$, and since the non-invariant terms are necessarily
$j$-dependent, we conclude that the invariance of the free fermion
Lagrangian can be established by the use of 
Eq.~(\ref{identity}) together with the cancellation of $j$-dependence.
This is the best we can do in proving invariance at the quaternionic
level. It is certainly interesting that the $j$-dependent terms we
systematically eliminate are also non-invariant, because it suggests
that a true quaternionic quantum mechanics may break symmetries that
are preserved by the complex subspace.

\section{Local Abelian transformations}

The possibility of independent phase rotations acting on the left and
right provides a natural motivation for the doubling of the gauge
symmetry. Indeed, local U(1)$\otimes$U(1) transformations are given by
\begin{equation}
\Psi_L(x) \to e^{i\alpha(x)} \Psi_L(x) e^{i\beta(x)}
\end{equation}
or, equivalently,
\begin{eqnarray}
\Psi_L(x) & \to & e^{i[\alpha(x) + \beta(x)]} \hat{C}_+ \Psi_L(x)
+ e^{i[\alpha(x) - \beta(x)]} \hat{C}_- \Psi_L(x) \nonumber\\
& = & e^{i[\alpha(x) + \beta(x)(\hat{C}_+ - \hat{C}_-)]} \Psi_L(x).
\label{locu1xu1}
\end{eqnarray}
The quaternionic covariant derivative is given by
\begin{equation}\label{d+A}
{\cal D}_{\mu} \Psi_L \equiv \partial_{\mu} \Psi_L + i \hat{\cal A}_{\mu}
\Psi_L,
\label{cov}
\end{equation}
where $\hat{\cal A}_{\mu}$ is the quaternionic gauge field with
gauge transformation law
\begin{equation}
\hat{\cal A}_{\mu} \to \hat{\cal A}_{\mu} - \partial_{\mu}\alpha -
\partial_{\mu}\beta (\hat{C}_+ - \hat{C}_-).
\label{xfm}
\end{equation}
It is interesting to display the detailed proof that the above really
leads to a covariant derivative. What we need to establish is that
${\cal D}_{\mu}\Psi_L$ transforms the same way as $\Psi_L$, namely
that
\begin{equation}
{\cal D}_{\mu}\Psi_L \to e^{i[\alpha + \beta(\hat{C}_+ - \hat{C}_-)]}
{\cal D}_{\mu}\Psi_L.
\end{equation}
Under a gauge transformation we have that
\begin{eqnarray}
{\cal D}_{\mu}\Psi_L & \to & e^{i[\alpha + \beta(\hat{C}_+ -
\hat{C}_-)]} \partial_{\mu}\Psi_L \nonumber \\
& + &i [\partial_{\mu}\alpha + \partial_{\mu}\beta(\hat{C}_+ -
\hat{C}_-)]e^{i[\alpha + \beta(\hat{C}_+ - \hat{C}_-)]}\Psi_L
\nonumber\\
& + & i[\hat{\cal A}_{\mu} - \partial_{\mu} \alpha - \partial_{\mu}\beta
(\hat{C}_+ - \hat{C}_-)] 
e^{i[\alpha + \beta(\hat{C}_+ - \hat{C}_-)]} \Psi_L\nonumber\\
& = & e^{i[\alpha + \beta(\hat{C}_+ -
\hat{C}_-)]} \partial_{\mu}\Psi_L + i \hat{\cal A}_{\mu} e^{i[\alpha +
\beta(\hat{C}_+ - \hat{C}_-)]} \Psi_L,
\end{eqnarray}
so the result we seek follows if $\hat{\cal A}_{\mu}$ commutes with the
exponential. Since
\begin{equation}
[\hat{C}_{\pm}, \hat{C}_+ - \hat{C}_-] = 0
\end{equation}
it follows that $\hat{\cal A}_{\mu}$ can be written as a linear
combination of the identity and $\hat{C}_{\pm}$ with vector fields as
coefficients. Since the identity is equal to $\hat{C}_+ + \hat{C}_-$,
we can without loss of generality write that 
\begin{equation}
\hat{\cal A}_{\mu} = g A_{\mu} \hat{C}_+ - g' A'_{\mu} \hat{C}_-,
\end{equation}
where the parameters $g$ and $g'$ are coupling constants, while 
$A_{\mu}$ and $A'_{\mu}$ are real gauge fields. Using Eq.~(\ref{xfm}),
we see that they have the transformation laws
\begin{equation}
A_{\mu} \to A_{\mu} - \frac{1}{g} \partial_{\mu}(\alpha + \beta)
\end{equation}
and
\begin{equation}
A'_{\mu} \to A'_{\mu} - \frac{1}{g'} \partial_{\mu}(-\alpha + \beta).
\end{equation}
The gauge invariant kinetic energy Lagrangian is
\begin{equation}
{\cal L}_{gK} = \frac{1}{2}( i \overline{\Psi}_L {\cal D}\!\!\!\!/ \Psi_L + 
\overline{\Psi}_L {\cal D}\!\!\!\!/ \Psi_L i ).
\end{equation}
The proof that this is gauge invariant follows exactly the same steps
as the 
proof that the free fermion Lagrangian is invariant under global
transformations, because we have a 
properly defined covariant derivative.

In symplectic components the quaternionic covariant derivative is
given by
\begin{equation}
{\cal D}_{\mu} \Psi_L = D_{\mu} \psi_L + j D'_{\mu} \psi'_R,
\end{equation}
and the kinetic energy Lagrangian is simply
\begin{equation}
{\cal L}_{gK} = i \overline{\psi}_L D\!\!\!\!/ \psi_L + i
\overline{\psi}'_R D\!\!\!\!/\,' \psi'_R,
\end{equation}
where
\begin{equation}
D_{\mu} \psi_L \equiv \partial_{\mu} \psi_L + i g A_{\mu} \psi_L\quad
{\rm and}\quad 
D'_{\mu} \psi'_R \equiv \partial_{\mu} \psi'_R + i g' A'_{\mu} \psi'_R.
\end{equation}
We will discuss the constraint due to parity invariance after we write
down the gauge boson kinetic energy terms in the next section.

\section{Gauge boson kinetic energy terms}

The gauge field $\hat{\cal A}_{\mu}$ is a natural object to use when
describing the coupling of gauge bosons to fermions, but it is not
so appropriate for the description of the gauge boson kinetic energy
terms. This is simply because $\hat{\cal A}_{\mu}$ depends on
$\hat{C}_{\pm}$ which have been defined to act on quaternionic fields.
Rather than trying to extend the definition of these operators to
include action on, for instance, state vectors, it is more convenient
to adopt the approach explained below.

Note that
\begin{equation}
(g A_{\mu} + j g' A'_{\mu}) (\hat{C}_+ + j\hat{C}_-)
= \hat{\cal A}_{\mu} + j (g A_{\mu} \hat{C}_- + g' A'_{\mu} \hat{C}_+),
\label{a}
\end{equation}
which allows us to rewrite the gauge invariant fermion kinetic energy
Lagrangian as
\begin{equation}
{\cal L}_{gK} = 
\frac{1}{2} \left[i\overline{\Psi}_L\left(\partial\!\!\!/ +
\frac{1}{2}(i{\cal A}\!\!\!/\,\hat{C} + {\cal A}\!\!\!/\:\hat{C}i)
\right) \Psi_L
+ \overline{\Psi}_L\left(\partial\!\!\!/ + 
\frac{1}{2}(i{\cal A}\!\!\!/\,\hat{C} + {\cal A}\!\!\!/\:\hat{C}i)
\right)\Psi_L i\right]
\end{equation}
where
\begin{equation}
{\cal A}_{\mu} \equiv g A_{\mu} + j g' A'_{\mu}
\end{equation}
and
\begin{equation}
\hat{C} \equiv \hat{C}_+ + j \hat{C}_-.
\end{equation}
This is true because the $j$ piece on the right-hand side of
Eq.~(\ref{a}) cancels between
the two terms in ${\cal L}_{gK}$. The field ${\cal A}_{\mu}$ is
clearly a natural object in a sense, because it is the gauge boson
analogue of $\Psi_L$. (Note that the symplectic components of ${\cal
A}_{\mu}$ are real fields rather than complex fields.)

The gauge boson kinetic energy terms are now easy to write down in a
quaternionic fashion. The field strength tensor ${\cal F}_{\mu\nu}$
is formed from ${\cal A}_{\mu}$ by applying the quaternionic covariant
derivative Eq.(\ref{d+A}),
\begin{equation}
{\cal F}_{\mu\nu} \equiv {\cal D}_{\mu} {\cal A}_{\nu} - 
{\cal D}_{\nu} {\cal A}_{\mu}.
\end{equation}
This is well defined, as the $\hat{C}_{\pm}$ act on quaternion valued
fields.  In terms of the symplectic components of ${\cal A}$,
\begin{eqnarray}
{\cal D}_{\mu} {\cal A}_{\nu} & = & (\partial_{\mu} + ig A_{\mu}\hat{C}_+
- ig' A'_{\mu}\hat{C}_- ) (g A_{\nu} + j g' A'_{\nu}) \\
& = & (g \partial_{\mu}A_{\nu} + ig^2 A_{\mu}A_{\nu}) + 
j(g' \partial_{\mu}A'_{\nu} + ig'^2 A'_{\mu}A'_{\nu}) \label{calF_A}. 
\end{eqnarray}
Hence, for our Abelian $A_{\mu}$ and $A'_{\mu}$ fields,
\begin{equation}
{\cal F}_{\mu\nu} = 
g (\partial_{\mu}A_{\nu} - \partial_{\nu}A_{\mu}) +
jg' (\partial_{\mu}A'_{\nu} - \partial_{\nu}A'_{\mu}),
\end{equation}
which is manifestly gauge invariant.

We define a complimentary gauge field by
\begin{equation}
{\cal A}_{\ast}^{\mu} \equiv g A^{\mu} - g' A'^{\mu} j.
\end{equation}
The corresponding gauge covariant derivative is
\begin{equation}
{\cal D}_{\ast}^{\mu} \equiv 
\partial^{\mu} + \hat{\cal A}_{\ast} .
\end{equation}
\begin{equation}
\hat{\cal A}_{\ast} \equiv i {\cal A}_{\ast}\hat{C} + {\cal A}_{\ast}\hat{C}i
= gA\hat{C}_+ - g'A'\hat{C}_-.
\end{equation}
Hence, the corresponding field strength tensor is
\begin{equation}
{\cal F}_{\ast}^{\mu\nu} \equiv 
{\cal D}_{\ast}^{\mu}{\cal A}_{\ast}^{\nu}
- {\cal D}_{\ast}^{\nu}{\cal A}_{\ast}^{\mu}
= g (\partial^{\mu}A^{\nu} - \partial^{\nu}A^{\mu}) 
- g' (\partial^{\mu}A'^{\nu} - \partial^{\nu}A'^{\mu})j .
\end{equation}

To form the  gauge boson kinetic energy Lagrangian, we observe that
\begin{equation} 
{\cal F}_{\ast}^{\mu\nu}{\cal F}_{\mu\nu} =
g^2 F^{\mu\nu}F_{\mu\nu} + g'^2 F'^{\mu\nu} F'_{\mu\nu}
+ j\;\mbox{term.}
\end{equation}
If parity symmetry is imposed, then $g=g'$. 
Hence, the gauge boson kinetic energy Lagrangian is
\begin{equation}
-\frac{1}{4} F_{\mu\nu} F^{\mu\nu} - \frac{1}{4} F'_{\mu\nu}
F'^{\mu\nu}
= \frac{-1}{8g^2}({\cal F}_{\ast}^{\mu\nu}{\cal F}_{\mu\nu}
- i {\cal F}_{\ast}^{\mu\nu}{\cal F}_{\mu\nu} i) .
\end{equation}

Our final task in this section is to write down the quaternionic
version of the gauge invariant kinetic energy mixing term between the
gauge fields: $F^{\mu\nu}F'_{\mu\nu}$.
This term is responsible for photon--mirror-photon and 
$Z$--mirror-$Z$ mixing in the EPM (note that it is parity invariant).
It is simply given by
\begin{equation}
F^{\mu\nu}F'_{\mu\nu} = \frac{-1}{8gg'} 
\left(i{\cal F}_{\ast}^{\mu\nu} ji {\cal F}_{\mu\nu}
+ {\cal F}_{\ast}^{\mu\nu} ji {\cal F}_{\mu\nu} i \right).
\end{equation}
This term is then multiplied by an arbitrary constant and added to the
diagonal kinetic energy terms for $A_{\mu}$ and $A'_{\mu}$. The
magnitude of photon--mirror-photon and $Z$--mirror-$Z$ mixing is
controlled by this parameter \cite{flv,glashow}. 

Before proceeding to non-Abelian transformations, we want to point out that 
the operator $\hat{C}$ is interesting in its own right. It belongs
to a class of idempotent operators given by $\hat{C}_+ + \exp(ia) j
\hat{C}_-$ where $a$ is an arbitrary phase. Its action on quaternionic
fields is simply
\begin{equation}
\hat{C} \Psi_R = \psi_R - \psi'_L.
\end{equation}
It therefore maps a chiral quaternionic spinor onto a Dirac-like
field. The field $\psi_R - \psi'_L$ is not a Dirac field in the usual
sense because $\psi_R$ and $-\psi'_L$ are not mass partners in the
EPM. However, it behaves as a Dirac field under Lorentz
transformations.
Just like $\hat{C}_{\pm}$, $\hat{C}$ projects one
representation of the Lorentz group onto another. Unlike
$\hat{C}_{\pm}$, $\hat{C}$ projects a reducible representation onto
another reducible representation. The arbitrary phase $a$ just
redefines the relative phases of the two chiral components of the
Dirac-like field. We will meet another of these operators when we
study Yukawa interactions.

\section{Non-Abelian transformations}

We now turn to non-Abelian gauge transformations. Suppose there are a
number of chiral quaternionic fermion fields which are placed into a
column matrix $\Psi_L$. In the Abelian case we were able in a natural
way to introduce two independent U(1) symmetries by left
multiplication and right multiplication. What is the analogue of this
for non-Abelian transformations? A sensible definition is obtained by
writing down the non-Abelian extension of Eq.~(\ref{locu1xu1}). The
operators $\hat{C}_{\pm}$ allow one to write group action from the
right as effectively group action from the left. The non-Abelian gauge
transformation of the column matrix $\Psi_L$ is then
\begin{equation}
\Psi_L \to e^{i \alpha^a T^a} 
e^{i \beta^b T^{b}(\hat{C}_+ - \hat{C}_-)} \Psi_L,
\label{locGxG}
\end{equation}
where the $T$'s are the generators of the appropriate representation
of an arbitrary group $G$, while the $\alpha$'s and $\beta$'s
are independent group parameters.
For convenience we will not combine the two exponentials into one,
although one could do so by using the Campbell-Baker-Hausdorff relation
(this was of course trivial in the Abelian case). To understand what
this transformation does, it is best to consider two special cases.
Consider first the subset of transformations defined by $\alpha^a =
\beta^a$. One then has
\begin{eqnarray}
\Psi_L & \to & e^{i \alpha^a T^a} 
e^{i \alpha^b T^{b}(\hat{C}_+ - \hat{C}_-)} \Psi_L\nonumber\\
& = & e^{i \alpha^a T^a} \left( e^{i \alpha^b T^b} \hat{C}_+ 
+ e^{- i\alpha^b T^b} \hat{C}_- \right) \Psi_L\nonumber\\
& = & e^{i 2\alpha^a T^a} \hat{C}_+ \Psi_L + \hat{C}_- \Psi_L,
\end{eqnarray}
which in symplectic components is simply
\begin{equation}
\psi_L \to e^{i 2\alpha^a T^a} \psi_L\quad {\rm and}\quad
\psi'_R \to \psi'_R.
\end{equation}
The complex left handed field transforms, but its mirror matter
partner does not. Now consider the subset of transformations defined
by $\alpha^a = -\beta^a$. This transformation is given by
\begin{eqnarray}
\Psi_L & \to & e^{i \alpha^a T^a}
e^{- i \alpha^b T^{b}(\hat{C}_+ - \hat{C}_-)} \Psi_L\nonumber\\
& = & e^{i \alpha^a T^a} \left( e^{- i \alpha^b T^b} \hat{C}_+ 
+ e^{i\alpha^b T^b} \hat{C}_- \right) \Psi_L\nonumber\\
& = & \hat{C}_+ \Psi_L + e^{i 2\alpha^a T^a} \hat{C}_- \Psi_L,
\end{eqnarray}
which in symplectic components now sees $\psi'_R$ transform and
$\psi_L$ remain invariant:
\begin{equation}
\psi_L \to \psi_L\quad {\rm and}\quad \psi'_R \to e^{-i 2\alpha^a
T^{a*}} \psi'_R.
\end{equation}
This shows that the transformations of Eq.~(\ref{locGxG}) belong to
the group $G\otimes G$, just as the Abelian transformations
considered earlier were members of U(1)$\otimes$U(1).

At first sight, it appears that if $\psi_L$ transforms as the $(R,1)$
representation of $G \otimes G$ then $\psi'_R$ must transform as the
$(1,R^*)$ representation. This is because $iT^a$ gets turned into
$-iT^{a*}$ after being commuted through $j$. While this is
perfectly true, it is equally valid to say that $\psi'_R$ transforms
under the $(1,R)$ representation. Any transformation in $R$ can be
obtained from an appropriate transformation in 
$R^*$ by redefining the group
parameters. Consider the fundamental representation of SU(N) 
for instance. The generators form two classes: those that are real
and those that are pure imaginary. In order to go from $R^*$ to $R$
one simply reverses the signs of all group parameters that multiply
the real generators. One can therefore always reinterpret 
$\exp(-i \alpha^a T^{a*})$ as $\exp( i \alpha'^a T^a)$ by relating
the $\alpha^a$ with $\alpha'^a$ in this way. Note that this does
{\it not} mean that $R$ and $R^*$ are necessarily 
equivalent representations in the usual sense of being
related to each other through a change of basis of the
representation space (similarity transformation). 

The gauge covariant derivative for non-Abelian $G\otimes G$
transformations can now be written down in close analogy to the
Abelian case described above. It is given by
\begin{equation}\label{d+W}
{\cal D}_{\mu} \Psi_L \equiv \partial_{\mu} \Psi_L 
+ i \hat{\cal W}_{\mu} \Psi_L. 
\end{equation}

In terms of the real components $W^a_{\mu}$ and $W'^b_{\mu}$,
the quaternionic gauge field is given by
\begin{equation}
\hat{\cal W}_{\mu} \equiv (g T^a W_{\mu}^a \hat{C}_+
- g' T^{\ast a}W'^{a}_{\mu} \hat{C}_-)\,,
\end{equation}
where $g$ and $g'$ are independent gauge coupling constants.

In terms of symplectic components, the covariant derivative of
$\Psi_L$ is given by
\begin{eqnarray}
{\cal D}_{\mu}\Psi_L & = & (\partial_{\mu} + i g W_{\mu}^a T^a)\psi_L +
j (\partial_{\mu} + i g' W'^{b}_{\mu} T^b)\psi'_R \nonumber\\
& \equiv & D_{\mu} \psi_L + j D'_{\mu} \psi'_R.
\end{eqnarray}

The gauge covariant kinetic energy term is
\begin{equation}
{\cal L}_{gK} = \frac{1}{2}
(i\overline{\Psi}_L {\cal D}\!\!\!\!/\ \Psi_L
+ \overline{\Psi}_L {\cal D}\!\!\!\!/\ \Psi_L i),
\end{equation}
which in terms of the symplectic components is simply
\begin{equation}
{\cal L}_{gK} = i \overline{\psi}_L D\!\!\!\!/\ \psi_L +
i \overline{\psi'}_R D\!\!\!\!/\ ' \psi'_R.
\end{equation}
This is precisely analogous to the Abelian case.

By demanding that the previous Lagrangian be gauge invariant, 
the non-Abelian gauge field $\hat{\cal W}_{\mu}$ must undergo
the gauge transformation
\begin{equation}
\hat{\cal W}_{\mu} \to 
U_{\alpha} U_{\beta} \hat{\cal W}_{\mu} U_{\beta}^{-1} 
U_{\alpha}^{-1} + i (\partial_{\mu} U_{\alpha} ) U_{\alpha}^{-1} 
+ i U_{\alpha}(\partial_{\mu} U_{\beta}) U_{\beta}^{-1}
U_{\alpha}^{-1}.
\end{equation}
where
\begin{equation}
U_{\alpha} \equiv e^{i\alpha^a T^a}
\quad{\rm and}\quad 
U_{\beta} \equiv e^{i\beta^b T^b(\hat{C}_+ - \hat{C}_-)},
\end{equation}
and $U_{\beta}^{-1} = U_{-\beta}$.

The infinitesimal version of this law is
\begin{equation}\label{dW}
\hat{\cal W}_{\mu} \to \hat{\cal W}_{\mu}
+ i [\alpha^a T^a, \hat{\cal W}_{\mu}]
+ i [\beta^b T^b (\hat{C}_+ - \hat{C}_-), \hat{\cal W}_{\mu}]
- \partial_{\mu}\alpha^a T^a
- \partial_{\mu}\beta^b T^b (\hat{C}_+ - \hat{C}_-).
\end{equation}

The analogy with the Abelian case is sufficiently strong to enable us to
form a quaternionic non-Abelian gauge field
\begin{equation}
{\cal W}^{\mu} \equiv (g T^a W_{\mu}^a + j g' T^a W'^{a}_{\mu})\,,
\end{equation}
and to rewrite the gauge invariant fermion kinetic Lagrangian as
\begin{equation}
{\cal L}_{gK} =
\frac{1}{2} \left(i\overline{\Psi}_L[\partial\!\!\!/ +
\frac{1}{2}(i{\cal W}\!\!\!\!/\,\hat{C} + {\cal W}\!\!\!\!/\,\hat{C}i)
]\Psi_L
+ \overline{\Psi}_L[\partial\!\!\!/ +
\frac{1}{2}(i{\cal W}\!\!\!\!/\,\hat{C} + {\cal W}\!\!\!\!/\,\hat{C}i)
]\Psi_L i\right).
\end{equation}

 From Eq.(\ref{dW}) we obtain the infinitesimal transformation laws for
the components:
\begin{equation}
W_{\mu}^a \to W_{\mu}^a +  f^{abc} (\alpha^c + \beta^c)
W_{\mu}^b - \frac{1}{g} \partial_{\mu}(\alpha^a + \beta^a)
\end{equation}
and
\begin{equation}
W'^{a}_{\mu} \to W'^{a}_{\mu} +  f^{abc} (\alpha^c - \beta^c)
W'^{b}_{\mu} + \frac{1}{g'} \partial_{\mu}(\alpha^a - \beta^a),
\end{equation}
where $f^{abc}$ are the real totally antisymmetric structure
constants of $G$.
This demonstrates that the $W_{\mu}^a$ are the gauge fields for one
of the factors in $G\otimes G$, while the $W'^{a}_{\mu}$ are the gauge 
fields for the other factor. 

A quaternionic derivation of the gauge invariance of ${\cal L}_{gK}$
can be given by close analogy with the proof in the Abelian case.
Since this is straightforward, we will not write down the details.

The field strength tensor for the quaternionic non-Abelian 
gauge field is formed by applying the quaternionic covariant
derivative Eq.(\ref{d+W}) to ${\cal W}$,
\begin{equation}
{\cal F}_{\mu\nu} \equiv 
{\cal D}_{\mu} {\cal W}_{\nu} - {\cal D}_{\nu} {\cal W}_{\mu}.
\end{equation}
In terms of the symplectic components of ${\cal W}$,
\begin{eqnarray}
{\cal D}_{\mu} {\cal W}_{\nu} & = & 
(\partial_{\mu} + ig T^b W^b_{\mu}\hat{C}_+
- ig' T^{\ast b}W'^b_{\mu}\hat{C}_-)
(g T^a W_{\mu}^a + j g' T^a W'^{a}_{\mu}) \nonumber\\
& = & (g T^a \partial_{\mu}W^a_{\nu} + ig^2 T^bT^a W^b_{\mu}W^a_{\nu}) +
j(g' T^a \partial_{\mu}W'_{\nu} + ig'^2 T^bT^a W'^b_{\mu}W'^a_{\nu}).
\nonumber\\
&&\label{calF_W}
\end{eqnarray}
Hence, for our non-Abelian $W^a_{\mu}$ and $W'^a_{\mu}$ fields,
\begin{eqnarray}
{\cal F}^a_{\mu\nu} &=&
g T^a(\partial_{\mu}W^a_{\nu} - \partial_{\nu}W^a_{\mu} 
- g f^{abc} W^b_{\mu}W^c_{\nu}) \nonumber\\
&+& jg' T^a(\partial_{\mu}W'^a_{\nu} - \partial_{\nu}W'^a_{\mu}
- g' f^{abc} W'^b_{\mu}W'^c_{\nu}).
\end{eqnarray}

The symplectic components of this expression are gauge invariant in
precisely the fashion we expect, indicating that ${\cal F}^a_{\mu\nu}$
is the natural object from which to construct the free field 
Lagrangian\footnote{
The gauge transformation law of its companion (operator valued)
field strength tensor is
$\hat{\cal F}_{\mu\nu} \to U_{\beta} U_{\alpha} \hat{\cal F}_{\mu\nu}
U_{\alpha}^{-1} U_{\beta}^{-1}$, as would be expected.}.
As for the Abelian case, in order to write down the gauge boson kinetic
energy terms a complimentary gauge field must be introduced,
\begin{equation}
{\cal W}_{\ast}^{\mu} \equiv (g T^a W^{a\mu} - g' T^a W'^{a\mu}j).
\end{equation}

Denoting the corresponding gauge covariant field strength by ${\cal F}_{\ast}$,
the gauge boson kinetic energy Lagrangian is
\begin{equation}
{\cal L}_{{\rm K}_{\mu}} =
\frac{-1}{8g^2}\left({\rm Tr}\{{\cal F}_{\ast}^{\mu\nu}{\cal F}_{\mu\nu}\}
- i\,{\rm Tr}\{{\cal F}_{\ast}^{\mu\nu}{\cal F}_{\mu\nu}\}\,i \right),
\end{equation}
where parity symmetry has been imposed, $g=g'$.
Note that gauge covariance prevents the appearance of a
kinetic energy mixing term in the non-Abelian case. The above
Lagrangian produces the standard expressions for the kinetic energy
terms for the real fields $W^a_{\mu}$ and $W'^a_{\mu}$.

\section{Higgs fields}

Let us now consider spin-0 fields. We will first determine the free
kinetic energy term, which we will then make gauge invariant. After
that we will write down the Higgs potential. Finally, we will examine
Yukawa couplings, which will complete our construction of the
quaternionic redrafting of the EPM.

The quaternionic Higgs field $\Phi$ is defined by analogy with $\Psi$:
\begin{equation}
\Phi \equiv \phi + j\phi',
\end{equation}
where $\phi$ and $\phi'$ are ordinary complex Higgs fields. Its free
kinetic energy term is simply
\begin{equation}
{\cal L}_{K\Phi} = \frac{1}{2}\left(
\partial^{\mu} \Phi^{\dagger} \partial_{\mu} \Phi
- i\partial^{\mu} \Phi^{\dagger} \partial_{\mu} \Phi i\right),
\end{equation}
which in symplectic components becomes
\begin{equation}
{\cal L}_{K\Phi} = \partial^{\mu} \phi^{\dagger} \partial_{\mu} \phi
+ \partial^{\mu} \phi'^{\dagger} \partial_{\mu} \phi'.
\end{equation}
The second term is needed to remove $j$-dependence from the Lagrangian.
One could also argue that it is not necessary, because the
$j$-dependent term from the first term is a 4-divergence and hence
would not contribute to the equations of motion even if it were there.
We will nevertheless remove it.

Both Abelian and non-Abelian gauge transformations are defined in
exactly the same way as for fermions. The gauge covariant derivatives
are also identical, leading to a gauge invariant Lagrangian which
simply sees $\partial_{\mu}$ replaced by the appropriate covariant
derivative. Since this is straightforward, we will not write down the
details.

We will now construct the Higgs potential for a Higgs multiplet $\Phi$
that transforms under a complex representation $(R,1) \oplus (1,R)$
of $G \otimes G$. We will further suppose that $R$ is chosen so that
$R\otimes R\otimes R$ and $R\otimes R\otimes R\otimes R$ do not
contain singlets. Therefore the only gauge singlet we need consider
for a renormalisable Higgs potential is that contained in $R^* \otimes
R$. These conditions simply reproduce those applicable to the Higgs
multiplet of the Standard Model, where $\phi^3$ and $\phi^4$ terms are
not gauge invariant.

In terms of the symplectic components $\phi$ and $\phi'$, the most
general Higgs potential is thus
\begin{equation}
V = -\mu^2 \phi^{\dagger}\phi - \mu'^2 \phi'^{\dagger}\phi'
+ \lambda (\phi^{\dagger}\phi)^2 
+ \lambda' (\phi'^{\dagger}\phi')^2
+ \kappa \phi^{\dagger}\phi \phi'^{\dagger}\phi'.
\label{V}
\end{equation} 
The relevant terms to consider for $\Phi$ are as follows:
\begin{eqnarray}
v_+ & \equiv & \frac{1}{2}\left(
\Phi^{\dagger}\Phi - i\Phi^{\dagger}\Phi i\right) 
 = \phi^{\dagger}\phi + \phi'^{\dagger}\phi';\\ \nonumber
v_- & \equiv & - \frac{1}{2}\left(i \Phi^{\dagger} i \Phi 
- \Phi^{\dagger} i \Phi i\right) =
\phi^{\dagger}\phi - \phi'^{\dagger}\phi'.
\end{eqnarray}
The most general renormalisable Higgs potential is thus
\begin{equation}
V = - \mu_+^2 v_+ - \mu_-^2 v_- + \lambda_+ v_+^2 + \lambda_- v_-^2 +
\lambda_{\pm} v_+ v_-,
\end{equation}
which is clearly equal to the Higgs potential in Eq.~(\ref{V}) using 
appropriate relations between the coefficients of the various
terms.

Under parity in the EPM, $\phi \leftrightarrow \phi'$. In the
quaternionic framework the involution 
\begin{equation}
\Phi \to j \Phi i,
\end{equation}
achieves this result up to a phase:
\begin{equation}
\phi + j\phi' \to -i\phi' + ji\phi.
\end{equation}

The terms $v_+$ and $v_-$ are thus
parity-even and parity-odd respectively, and the Higgs potential of the
EPM is 
\begin{equation}
V_{\rm EPM} = - \mu_+^2 v_+ + \lambda_+ v_+^2 + \lambda_- v_-^2.
\end{equation}
An equivalent, and more convenient form is
\begin{equation}
V_{\rm EPM} = \kappa_+ ( v_+ - 2u^2 )^2 + \kappa_- v_-^2.
\end{equation}
This is precisely the Higgs potential first written down in 
Ref.\cite{flv} for the EPM. (By way of reminder, 
in the parameter space region $\kappa_{\pm} > 0$, the Higgs
potential is minimised by setting $\langle\phi\rangle =
\langle\phi'\rangle = u$ which is the parity conserving breakdown
pattern we want.)

\section{Yukawa interactions}

Finally, let us consider Yukawa couplings. We will restrict ourselves
to reproducing the Yukawa Lagrangian found in the EPM. A typical term
in the Lagrangian
is $\overline{q}_L \phi u_R + \overline{q}'_R \phi' u'_L$, where $q_L$
is the left-handed quark doublet, $u_R$ the right-handed up quark,
$\phi$ the Higgs doublet, and the primed fields are the mirror matter
partners of the standard fields. In terms of our notation above, we
therefore seek a quaternionic reformulation of the Yukawa pattern
$\overline{\psi}_L \phi \psi_R + \overline{\psi}'_R \phi' \psi'_L$.
This will allow us to write down the entire Yukawa Lagrangian of the
EPM.

The required quaternionic reformulation is
\begin{equation}
{\cal L}_{\rm Yuk} = \frac{h}{2}\left(i\overline{\Psi}_L
\frac{1}{2}(\Phi\hat{C}_{\ast} -i\Phi\hat{C}_{\ast}i)\Psi_R
+ \overline{\Psi}_L
\frac{1}{2}(\Phi\hat{C}_{\ast} -i\Phi\hat{C}_{\ast}i)
\Psi_Ri\right),
\end{equation}
where
\begin{equation}
\hat{C}_{\ast} \equiv \hat{C}_+ - j \hat{C}_-
\end{equation}
and $h$ is the Yukawa coupling constant. The symmetric treatment of
multiplication by $i$ is again required to remove $j$-dependence from
the Lagrangian. It is easy to check that this Lagrangian yields the
parity symmetric combination quoted in the previous paragraph. (The 
parity odd combination is obtained by inserting an additional factor of
$i$ between $\overline{\Psi}_L$ and $\Psi_R$.) 

\section{Exact Parity Model}

By applying all of the techniques developed above, the Lagrangian of
the Exact Parity Model can be rewritten in terms of quaternionic
fields. We sketch the outline of this below.

The EPM has gauge group $G_{SM}\otimes G_{SM}$ under which 
a generation of standard fermions has the multiplet structure
\[
q_L \sim (3,2,1/3;1,1,0),\quad u_R \sim (3,1,4/3;1,1,0),\quad
d_R \sim (3,1,-2/3;1,1,0),
\]
\begin{equation}
f_L \sim (1,2,-1;1,1,0),\quad e_R \sim (1,1,-2;1,1,0),
\end{equation}
while a generation of mirror fermions is given by
\[
q'_R \sim (1,1,0;3,2,1/3),\quad u'_L \sim (1,1,0;3,1,4/3),\quad
d'_L \sim (1,1,0;3,1,-2/3),
\]
\begin{equation}
f'_R \sim (1,1,0;1,2,1),\quad e'_L \sim (1,1,0;1,1,-2).
\end{equation}
Ordinary and mirror pairs can now be written as the symplectic
components of chiral quaternionic spinors:
\begin{eqnarray}
& Q_L = q_L + j q'_R,\quad U_R = u_R + j u'_L,\quad 
D_R = d_R + j d'_L,& \nonumber\\ 
& F_L = f_L + j f'_R,\quad E_R = e_R + j e'_L.&\ 
\end{eqnarray}
The $G_{SM}\otimes G_{SM}$ gauge transformations are then defined as
explained in Sections 5 and 7 above. Similarly, the gauge boson and
mirror gauge boson fields are assembled into quaternionic fields as
per Sections 6 and 7 and their kinetic energy terms are written down.

We next introduce the Higgs doublet $\phi$ and its mirror partner
$\phi'$,
\begin{equation} 
\phi \sim (1,2,1;1,1,0)\quad {\rm and}\quad \phi' \sim (1,1,0;1,2,1)
\end{equation}
which we then incorporate into a quaternionic scalar field $\Phi$:
\begin{equation}
\Phi = \phi + j \phi'.
\end{equation}
The Higgs boson gauge invariant kinetic energy terms and the Higgs
potential are then written down exactly as in Section 8. Yukawa
interactions are introduced as is Section 9, with the constraints of
gauge and parity invariance imposed. 
After spontaneous electroweak symmetry
breakdown, the Yukawa interactions lead to fermion mass terms of the
form given by ${\cal B}_5$ in Section 3. 

Parity symmetry is of course imposed by demanding invariance under
\begin{equation}
Q_L \to j \gamma^0 Q_L,\quad U_R \to j \gamma^0 U_R,\quad etc.
\end{equation}
for the fermion fields and
\begin{equation}
\Phi \to j \Phi i
\end{equation}
for the Higgs field. The gauge boson field parity transformation is,
\footnote{Again, only up to phases: 
$W^{\mu} + j W'^{\mu} \to iW'_{\mu} - j i W_{\mu}$.}
\begin{equation}
{\cal W}^{\mu} \to -j {\cal W}_{\mu}i,
\end{equation}
where ${\cal W}^{\mu}$ represents either gluons or the electroweak 
bosons. These parity transformations cause corresponding
Yukawa and gauge coupling constants in the two sectors to be equal.

This completes the construction of the minimal EPM. Other work has
shown that nonzero neutrino masses are desirable in the EPM because
they can naturally explain the solar and atmospheric neutrino anomalies
\cite{nu}. 
The easiest way to do this is to introduce a right-handed neutrino field
$\nu_R$ and its mirror partner $\nu'_L$. These constitute another
chiral quaternionic spinor $N_R$ where
\begin{equation}
N_R = \nu_R + j \nu'_L.
\end{equation}
Being a gauge singlet, $N_R$ can have bare mass in addition to the
electroweak mass it gains because of spontaneous symmetry breaking.
Two types of bare mass are possible. The first is of the ${\cal B}_1$
form and the second is of the ${\cal B}_3$ form, leading to the bare
mass Lagrangian
\begin{equation}
{\cal L}_{\rm mass} = \frac{m'}{2} (\overline{N}_R j N_R - i
\overline{N}_R j N_R i) + \frac{M}{2}\left[\overline{N}_R (N_R)^c
- i \overline{N}_R (N_R)^c i + {\rm H.c.}\right].
\end{equation}
The mass $m'$ mixes ordinary and mirror neutrinos via a
$\overline{\nu}_R\nu'_L$ term, while $M$ is a common Majorana mass for
$\nu_R$ and $\nu'_L$. The see-saw mechanism for explaining why the
observed neutrinos are exceptionally light is invoked by requiring
that $M$ be much larger than any other type of neutrino mass.

\section{Conclusion}

We conclude with some philosophical or interpretative remarks: 
What we have achieved in this paper is a quaternionic reformulation of
the EPM, but one explicitly based on standard complex quantum
mechanics. We have deliberately constructed our Lagrangian so that all
of the $j$- and $k$-dependent terms cancel out when the Lagrangian is
decomposed into its constituent real and complex fields. A more
profound goal would have been to incorporate the EPM into a quantal
structure that was explicitly quaternionic.
Such a program would be premature, however, because 
in the absence of empirical evidence 
there is no obviously best way to formulate quaternionic quantum mechanics.
Nevertheless, the above analysis is interesting in that it
demonstrates a connection between a specific algebraic structure (the
quaternions) and the model-building idea that the gauge group of the
world is a product of two isomorphic factors. This idea was originally
motivated for quite different reasons, such as to reinstate parity as
an exact symmetry \cite{lee,flv} or, 
more recently, to explain the observed neutrino anomalies \cite{nu}.
The present paper provides yet another reason to be
interested in $G_{SM} \otimes G_{SM}$ gauge theory.

\end{document}